\documentclass[runningheads]{llncs}

\usepackage[T1]{fontenc}
\usepackage{graphicx}
\usepackage{amsmath,amssymb,amsfonts}
\usepackage{booktabs}
\usepackage{xcolor}
\usepackage{enumitem}
\usepackage{mathtools}
\usepackage{tikz}
\usetikzlibrary{arrows.meta,positioning,fit,backgrounds,shapes.misc}

\definecolor{PaperBlue}{HTML}{2563EB}
\definecolor{PaperBlueFill}{HTML}{EAF2FF}
\definecolor{PaperOrange}{HTML}{D97706}
\definecolor{PaperOrangeFill}{HTML}{FFF1E5}
\definecolor{PaperGray}{HTML}{6B7280}
\definecolor{PaperGrayFill}{HTML}{F3F4F6}
\definecolor{PaperGrayBorder}{HTML}{D1D5DB}

\newcommand{\protoK}{Kademlia}
\newcommand{\protoCV}{Cyclon+Vicinity}
\newcommand{\udelta}{U_\Delta}

\begin{document}

\title{Usable Agent Discovery for Decentralized AI Systems}

\titlerunning{Usable Agent Discovery for Decentralized AI Systems}

\author{Patrizio Dazzi\inst{1,2} \and Emanuele Carlini\inst{2} \and Matteo Mordacchini\inst{3} \and Saul Urso\inst{1,2}}
\authorrunning{P. Dazzi et al.}

\institute{
Department of Computer Science \\ University of Pisa \\ \email{patrizio.dazzi@unipi.it}, \email{saul.urso@phd.unipi.it}
\and
Institute of Information Science and Technologies \\ National Research Council of Italy \\ \email{emanuele.carlini@isti.cnr.it}
\and
Institute of Informatics and Telematics \\ National Research Council of Italy \\ \email{matteo.mordacchini@iit.cnr.it}
}

\maketitle

\begin{abstract}
Large-scale agentic systems run on distributed infrastructures where many software agents share physical hosts and are discovered via peer-to-peer mechanisms. Discovery must handle node-level churn from failures and host departures and agent-level churn from demand-driven activation, deactivation, and state changes. Their interaction reshapes classic trade-offs between structured and unstructured overlays.
We study decentralized agent discovery under this two-level churn, assuming nodes host multiple agents, overlays are structured or gossip-based, and agents switch between warm and cold states. Using \protoK{} as a structured and \protoCV{} as a gossip baseline, we compare stable, node-churn-only, agent-cooling-only, and combined regimes to see when routing efficiency, resilience, and service readiness align or favor different designs.
Structured overlays are more robust and efficient in stable and node-churn regimes, while gossip-based overlays remain competitive and can be faster when readiness dominates.

\end{abstract}

\section{Introduction}
Large-scale agentic systems increasingly run on distributed infrastructures where many software agents share hosts and are discovered via decentralized mechanisms~\cite{dazzi2025iaia,carlini2023smartorc}. Following recent approaches such as AGNTCY~\cite{muscariello2025agntcy}, we assume agents are defined by the {\em skills} (capabilities) they expose and that discovery queries target these skills. Multiple agents may expose overlapping skills.
Discovery must locate not just the right node but a \emph{usable} execution target: a host may be reachable and correctly referenced while the requested agent is cold, suspended, or otherwise unable to serve due to a demand-driven lifecycle.
This complicates the usual comparison between structured and unstructured discovery substrates. Structured overlays~\cite{stoica2001chord,ratnasamy2001can,ferrucci2016multidimensional} provide low-latency lookups and compact routing state when membership is stable and routing invariants hold. Gossip-based (unstructured) overlays~\cite{jelasity2007gossip} incur higher messaging overhead but tolerate weaker consistency and often degrade more gracefully under change. For agentic infrastructures, this dichotomy is incomplete unless we separate two instability sources. At the infrastructure level, \emph{node-level churn} (failures, departures, recoveries) changes overlay connectivity and affects all agents on a node. At the service level, \emph{agent-level churn} arises from demand-driven lifecycles: agents cool after inactivity and must be reactivated on demand. These processes stress discovery differently and may favor different designs.
We therefore study decentralized agent discovery experimentally across four regimes: stable operation, node-level churn only, agent-level cooling only, and both combined. Our goal is not to crown a universally superior overlay, but to identify when structured and gossip-based approaches are favored by operating conditions. We compare a structured baseline based on \protoK{}~\cite{maymounkov2002kademlia} (the indexing substrate used in AGNTCY) with a gossip-based baseline based on \protoCV{}~\cite{voulgaris2005epidemic}, a representative neighborhood-maintenance scheme, extending our prior work on P2P discovery~\cite{carlini2009srds,baraglia2011group}.
Our contributions are: (i) a system model with agents' warm/cold states, (ii) observables separating efficiency, resilience, and readiness, and (iii) an empirical regime map for the two overlay families.

\section{Related Work}
\label{sec:related}

Structured peer-to-peer overlays such as Chord, CAN, and Kademlia provide efficient lookup by maintaining explicit routing invariants over participating nodes~\cite{stoica2001chord,ratnasamy2001can,maymounkov2002kademlia}. Their strength lies in predictable hop counts and compact state when membership is stable. Unstructured approaches based on flooding, peer sampling, or gossip trade that predictability for resilience, degrading more gracefully under churn at the cost of higher messaging overhead~\cite{demers1987epidemic,jelasity2007gossip,eugster2004epidemic}. This tension has shaped decades of decentralized systems design, yet it assumes that discovery targets are nodes themselves.
Multi-agent infrastructures break that assumption. Discovery must return not only a reachable host but a \emph{usable} execution context, because many agents co-reside on the same node and their runtime state evolves independently of the overlay. Our earlier work on decentralized service and resource discovery already highlighted how capability-based queries interact with P2P substrates~\cite{carlini2009srds}, but it did not separate node dynamics from service readiness.
The second dimension — readiness — connects directly to warm and cold execution in serverless platforms, where startup latency becomes part of the end-to-end cost~\cite{jonas2019serverless}. In agent systems, cooling is not an exception but a lifecycle policy, and it interacts with overlay maintenance: a correct route to a cold agent is still a failed request from the user's perspective. This coupling has received limited attention in the P2P literature.
A complementary line of work has explored gossip-based self-organization for semantic affinity and approximate matching. Prior results on community formation and recommender overlays showed how unstructured neighborhoods can be shaped toward skill similarity without global indexes~\cite{baraglia2011group,dazzi2011approximate,baraglia2013recommender}. More recently, the Internet of AI Agents perspective and continuum-oriented orchestration work have reinforced the need to evaluate discovery jointly with service-level objectives and distributed operating conditions~\cite{dazzi2025iaia,carlini2023smartorc,ferrucci2024replica}.
What remains open — and what this paper addresses — is the interaction of these two levels of churn. Existing comparisons treat node instability and agent lifecycle as separate concerns; we study them together, and we introduce a {\em useful availability} metric $\udelta$ to capture whether a discovered route yields a service within a meaningful deadline. This shifts the evaluation from "which overlay is faster" to "under which regime does robustness convert into usability."
\section{System Model}
\label{sec:model}

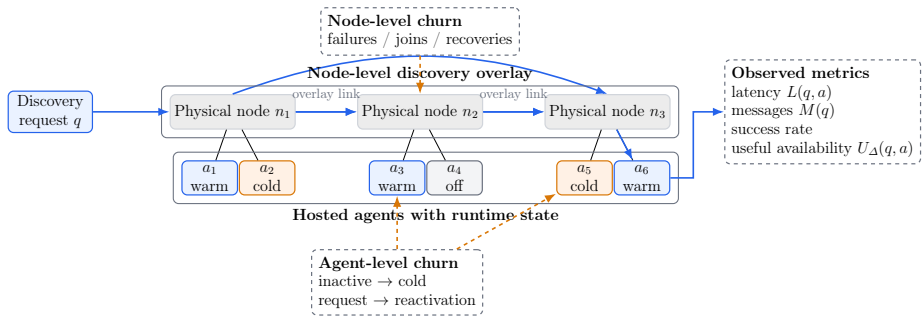
\begin{figure}[t]
\centering
\resizebox{\textwidth}{!}{%
\begin{tikzpicture}[
    font=\large,
    >=Latex,
    node distance=0.8cm and 1.0cm,
    box/.style={
        draw=PaperGrayBorder,
        fill=PaperGrayFill,
        rounded corners,
        align=center,
        line width=0.9pt,
        minimum width=2.2cm,
        minimum height=0.9cm
    },
    agentwarm/.style={
        draw=PaperBlue,
        fill=PaperBlueFill,
        rounded corners,
        align=center,
        line width=0.9pt,
        minimum width=1.45cm,
        minimum height=0.75cm
    },
    agentcold/.style={
        draw=PaperOrange,
        fill=PaperOrangeFill,
        rounded corners,
        align=center,
        line width=0.9pt,
        minimum width=1.45cm,
        minimum height=0.75cm
    },
    agentoff/.style={
        draw=PaperGray,
        fill=PaperGrayFill,
        rounded corners,
        align=center,
        line width=0.9pt,
        minimum width=1.45cm,
        minimum height=0.75cm
    },
    note/.style={
        draw=PaperGray,
        fill=white,
        dashed,
        rounded corners,
        align=left,
        inner sep=5pt,
        line width=0.9pt
    },
    flow/.style={->, very thick, draw=PaperBlue},
    churn/.style={->, dashed, very thick, draw=PaperOrange},
    every fit/.style={draw=PaperGray, rounded corners, inner sep=6pt, line width=0.8pt}
]

\node[box, draw=PaperBlue, fill=PaperBlueFill] (query) {Discovery\\request $q$};

\node[box, fill=gray!15, right=2.0cm of query] (n1) {Physical node $n_1$};
\node[box, fill=gray!15, right=1.6cm of n1] (n2) {Physical node $n_2$};
\node[box, fill=gray!15, right=1.6cm of n2] (n3) {Physical node $n_3$};

\draw[flow] (n1) -- node[above, yshift=5pt, text=PaperGray] {\normalsize overlay link} (n2);
\draw[flow] (n2) -- node[above, yshift=5pt, text=PaperGray] {\normalsize overlay link} (n3);
\draw[flow, bend left=20] (n1.north) to node[above, yshift=4pt, text=PaperGray] {\normalsize structured / unstructured} (n3.north);

\node[agentwarm, below=0.8cm of n1, xshift=-0.6cm] (a11) {$a_1$\\warm};
\node[agentcold, below=0.8cm of n1, xshift=0.9cm] (a12) {$a_2$\\cold};

\node[agentwarm, below=0.8cm of n2, xshift=-0.6cm] (a21) {$a_3$\\warm};
\node[agentoff, below=0.8cm of n2, xshift=0.9cm] (a22) {$a_4$\\off};

\node[agentcold, below=0.8cm of n3, xshift=-0.6cm] (a31) {$a_5$\\cold};
\node[agentwarm, below=0.8cm of n3, xshift=0.9cm] (a32) {$a_6$\\warm};

\draw[thin] (n1) -- (a11);
\draw[thin] (n1) -- (a12);
\draw[thin] (n2) -- (a21);
\draw[thin] (n2) -- (a22);
\draw[thin] (n3) -- (a31);
\draw[thin] (n3) -- (a32);

\node[fit=(n1)(n2)(n3), label=above:{\bfseries Node-level discovery overlay}] {};
\node[fit=(a11)(a12)(a21)(a22)(a31)(a32), label=below:{\bfseries Hosted agents with runtime state}] {};

\draw[flow] (query) -- (n1);
\draw[flow] (n1) -- (n2);
\draw[flow] (n2) -- (n3);
\draw[flow] (n3) -- (a32);

\node[note, above=1.0cm of n2] (nodechurn) {\textbf{Node-level churn}\\failures / joins / recoveries};
\draw[churn] (nodechurn) -- (n2);

\node[note, below=1.4cm of a21] (agentchurn) {\textbf{Agent-level churn}\\inactive $\rightarrow$ cold\\request $\rightarrow$ reactivation};
\draw[churn] (agentchurn) -- (a21);
\draw[churn] (agentchurn) -- (a31);

\node[note, right=1.4cm of n3] (metrics) {
\textbf{Observed metrics}\\
latency $L(q,a)$\\
messages $M(q)$\\
success rate\\
useful availability $\udelta(q,a)$
};
\draw[flow] (a32.east) -- ++(0.6,0) |- (metrics.west);

\end{tikzpicture}
}
\caption{Conceptual model of decentralized agent discovery under two-level churn.}
\label{fig:conceptual-model}
\end{figure}

In the following, we present the model depicted in Figure~\ref{fig:conceptual-model}. Specifically, we consider a time-varying set of physical nodes $N(t)$ connected by an overlay graph $G(t)$, and a set of hosted agents $A(t)$. Each agent $a \in A(t)$ is mapped to a host node, and multiple agents may be co-located on the same node.
Each agent has a runtime state
\[
\sigma(t,a) \in \{\textsc{warm}, \textsc{cold}, \textsc{off}\},
\]
where \textsc{warm} agents can serve immediately, \textsc{cold} agents are discoverable but incur startup delay, and \textsc{off} agents are unavailable, although they may still be known to the system through stale information.
We distinguish two sources of dynamics. Node-level churn affects $N(t)$ and $G(t)$ through joins, failures, and recoveries, potentially removing or reconnecting multiple agents at once. Agent-level churn affects $\sigma(t,a)$ through lifecycle transitions such as cooling after inactivity and later reactivation. These two processes are coupled but not equivalent: node failures remove sets of agents, whereas agent-level changes affect service readiness while the host remains reachable.
A discovery request $q$ is issued by a source node $s$ at time $t$ and targets a capability (or skill) $\kappa$. Discovery returns an \emph{agent record} identifying a candidate agent $a$ together with its host node. In the system model considered here, requests are skill-based and each skill may be replicated across multiple agents.
The end-to-end latency associated with serving $q$ via $a$ is
\[
L(q,a) = L_{\text{disc}}(q) + L_{\text{route}}(q,a) + L_{\text{start}}(a),
\]
where $L_{\text{disc}}(q)$ is discovery time, $L_{\text{route}}(q,a)$ is the time to reach the host node, and $L_{\text{start}}(a)$ is the startup delay, non-zero only when $\sigma(t,a)=\textsc{cold}$.
We distinguish between discovery success and service usability. A request is successful if discovery returns an agent $a$ and its host node is reachable:
\[
S(q,a) = \mathbf{1}\!\left[\text{$a$ is discovered and its host is reachable}\right]
\]
Useful availability captures whether the discovered agent can serve within a latency budget:
\[
\udelta(q,a) = \mathbb{E}\big[u_\Delta(L(q,a))\big],
\]
where $u_\Delta(x)=1$ if $x \le \Delta$ and $u_\Delta(x)=0$ otherwise. In practice, the expectation is estimated over the workload and the churn processes. Thus, success captures routing robustness, whereas $U_\Delta$ captures service usability.
We measure communication cost directly from the message traffic generated by each workload. In the analysis, we distinguish request-plane traffic from background maintenance traffic and use them together with $L$ and $\udelta$ to evaluate efficiency and usability.

\section{Research Questions and Expected Regimes}
\label{sec:goals}

The study asks a simple but consequential question: \emph{how does the choice for the creation and maintenance of the overlay graph $G(t)$ affect decentralized agent discovery once both physical hosts and hosted agents become dynamic?}

More specifically, we separate four regimes that are often conflated in practice: i) stable infrastructures with mostly warm agents, ii) infrastructures stressed mainly by node churn, iii) infrastructures stressed mainly by agent cooling, and iv) regimes in which the two forms of instability interact.
This leads to four research questions: under favorable conditions, does a structured overlay translate routing efficiency into better end-to-end behavior, or mainly into lower message cost? As node-level churn increases, does the structured advantage erode because repair and stale state begin to dominate lookup efficiency? When agents cool aggressively after inactivity, does precise host discovery become less important than readiness and cold-start delay? Under fully combined two-level churn, is there a regime in which the unstructured alternative offers better service usability despite higher communication overhead?

\section{Experimental Design and Evaluation Framework}
\label{sec:experiments}

The evaluation uses an event-driven SimPy simulator where physical nodes host multiple agents and join either a structured or unstructured overlay. For the structured overlay we use \protoK{}, the same indexing substrate as AGNTCY; for the unstructured overlay we use a \protoCV{} scheme, representing gossip-based neighborhood construction where peer sampling and local similarity jointly shape discovery. We vary five factor families: overlay type, node-level churn, agent cooling policy, cold-start cost, and workload structure. Runs use multiple random seeds and are summarized with confidence intervals. Main observables are end-to-end latency, message overhead, success rate, and useful availability $\udelta$. Success rate is the empirical fraction of requests for which $S(q,a)=1$.

The experimental plan is incremental: we establish a stable baseline, then add node-level churn, then agent cooling, and finally combine both into a joint instability ladder. This order is needed so the combined effects can be interpreted from already-understood individual effects. Workload sensitivity is examined separately on selected operating points as a second-stage extension.

Unless noted otherwise, synthetic experiments use 4096 logical agents on 2048 physical hosts (two agents per host) and a catalog of 50 skills. Each agent exposes one skill; host profiles are the union of hosted agent cards; requests originate from internal nodes; and discovery returns only the top candidate (\texttt{top\_k}=1). The substrate network is an Erdős–Rényi graph with zero packet loss and unit base latency. \protoK{} uses replication 3, $\alpha=3$, bucket size 8, publish TTL 60, and republish period 20. \protoCV{} uses logarithmic request TTL, Cyclon cache size 20, Cyclon shuffle length 5, Vicinity cache size 10, and shuffle period 20. The main instability ladders run for 80 simulated cycles with five repetitions, request rate 0.15, and warmup 25; the stable baseline runs for 60 cycles with five repetitions, request rate 0.125, and warmup 20. Table~\ref{tab:main-regimes} reports the regime-specific churn and lifecycle parameters used by the main figures.

\subsection{Experimental Regimes}

\textbf{\textit{E1: stable baseline.}}
This family establishes the reference regime under low infrastructure churn and long warm retention. Its role is to reveal the best-case trade-off between structured efficiency and unstructured redundancy. In our current implementation, E1 also supports a reduced-scale semantic ablation in which the synthetic task vocabulary and the skill range per agent are varied.

\noindent\textbf{\textit{E2: node-level churn only.}}
Here physical node failures and recoveries are increased progressively while agent cooling remains disabled or negligible. This family isolates overlay maintenance, stale routing, and disruption caused by host instability.

\noindent\textbf{\textit{E3: agent-level cooling only.}}
In this family the infrastructure remains stable, but agents cool aggressively after inactivity. The goal is to measure when cold-start penalties dominate end-to-end usability and weaken the practical benefit of accurate lookup.

\noindent\textbf{\textit{E4: combined two-level churn.}}
This is the central family of the paper. Node failures and agent cooling are activated together through a ladder from favorable to critical regimes. The purpose is to expose non-linear interactions and possible regime changes in which the preferred overlay type shifts.

\noindent\textbf{\textit{E5: workload sensitivity.}}
Workload skew and temporal locality are treated as a later sensitivity layer. They matter for warmness policies and discovery locality, but they should be explored only after the four core families above are stable and interpretable.

At this stage of the study, the core multi-host ladders summarized in
Table~\ref{tab:main-regimes} are complete and are read through five
complementary views: a latency/overhead trade-off, an explicit maintenance-cost
view, a degradation ladder across regimes, direct useful-availability curves,
and a reduced-scale semantic ablation. Together these views are enough to
compare both the main operating points and the way semantic sparsity changes the
comparison.

\begin{table}[t]
\centering
\small
\setlength{\tabcolsep}{4pt}
\caption{Main operating regimes used in Figures~\ref{fig:latency-overhead-tradeoff}--\ref{fig:useful-availability}. Session, downtime, ready, suspended, and warming values are means in simulated cycles; ``--'' is a disabled process.}
\label{tab:main-regimes}
\begin{tabular}{lrrrrr}
\toprule
Regime & Session & Downtime & Ready & Suspended & Warming \\
\midrule
Stable & -- & -- & -- & -- & -- \\
Node churn mild & 180 & 10 & -- & -- & -- \\
Node churn moderate & 100 & 30 & -- & -- & -- \\
Node churn aggressive & 60 & 45 & -- & -- & -- \\
Cooling mild & -- & -- & 60 & 3 & 1 \\
Cooling moderate & -- & -- & 20 & 6 & 2 \\
Cooling aggressive & -- & -- & 8 & 10 & 4 \\
Combined mild & 180 & 10 & 60 & 3 & 1 \\
Combined moderate & 100 & 30 & 20 & 6 & 2 \\
Combined critical & 100 & 30 & 8 & 10 & 4 \\
\bottomrule
\end{tabular}
\end{table}

\begin{figure}[t]
\centering
\includegraphics[width=\textwidth]{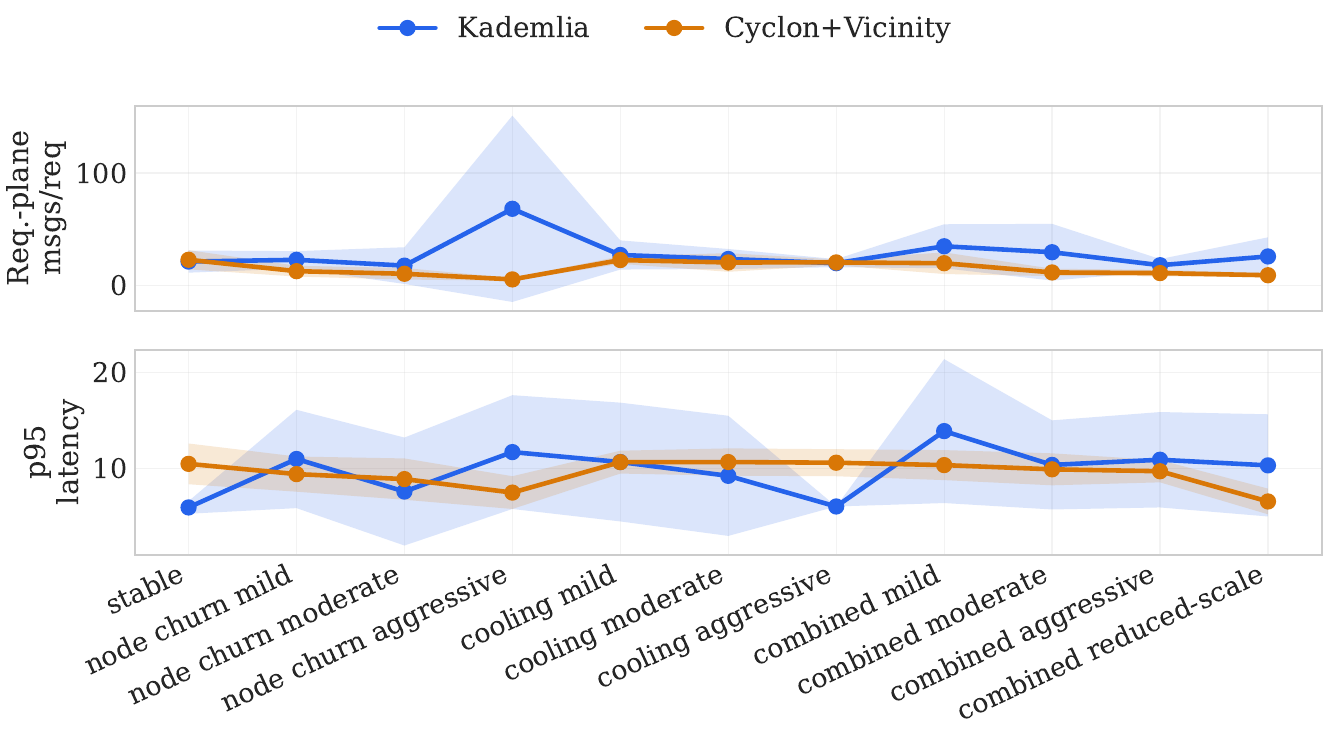}
\caption{Latency/overhead trade-off across the main operating regimes.}
\label{fig:latency-overhead-tradeoff}
\end{figure}

\begin{figure}[t]
\centering
\includegraphics[width=\textwidth]{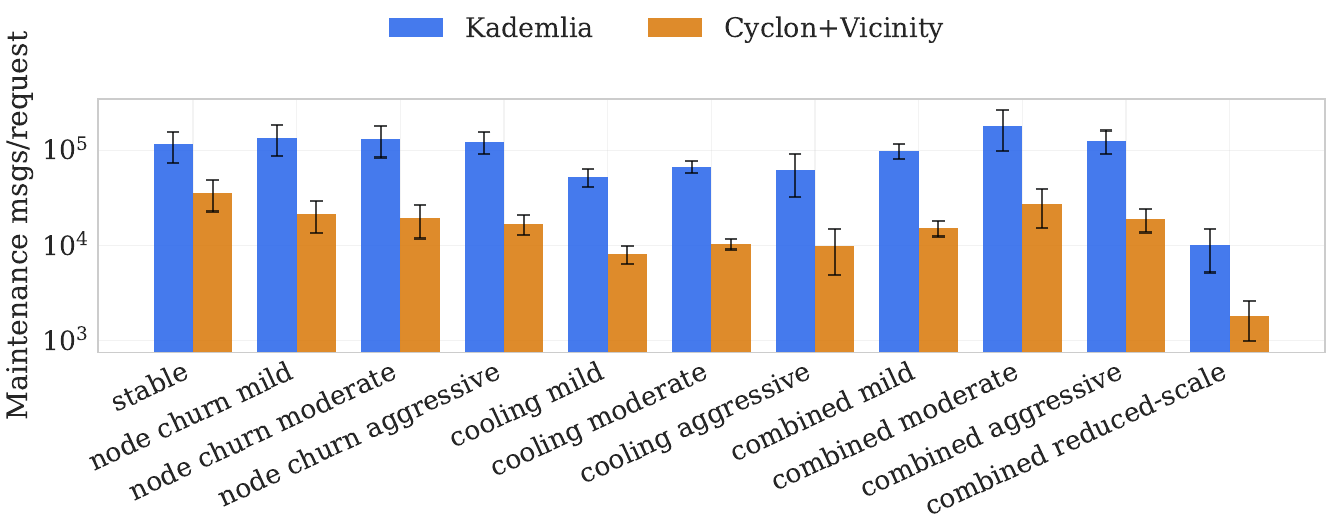}
\caption{Maintenance vs request-plane cost.}
\label{fig:maintenance-cost}
\end{figure}

\begin{figure}[t]
\centering
\includegraphics[width=\textwidth]{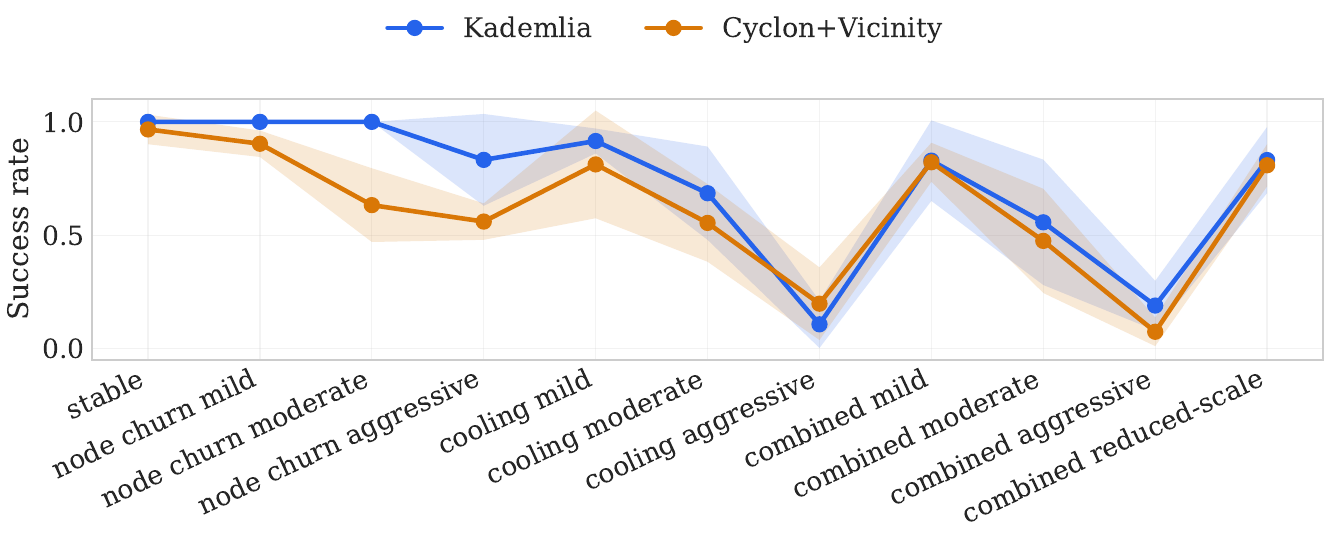}
\caption{Success rate across regimes.}
\label{fig:success-across-regimes}
\end{figure}

\begin{figure}[t]
\centering
\includegraphics[width=\textwidth]{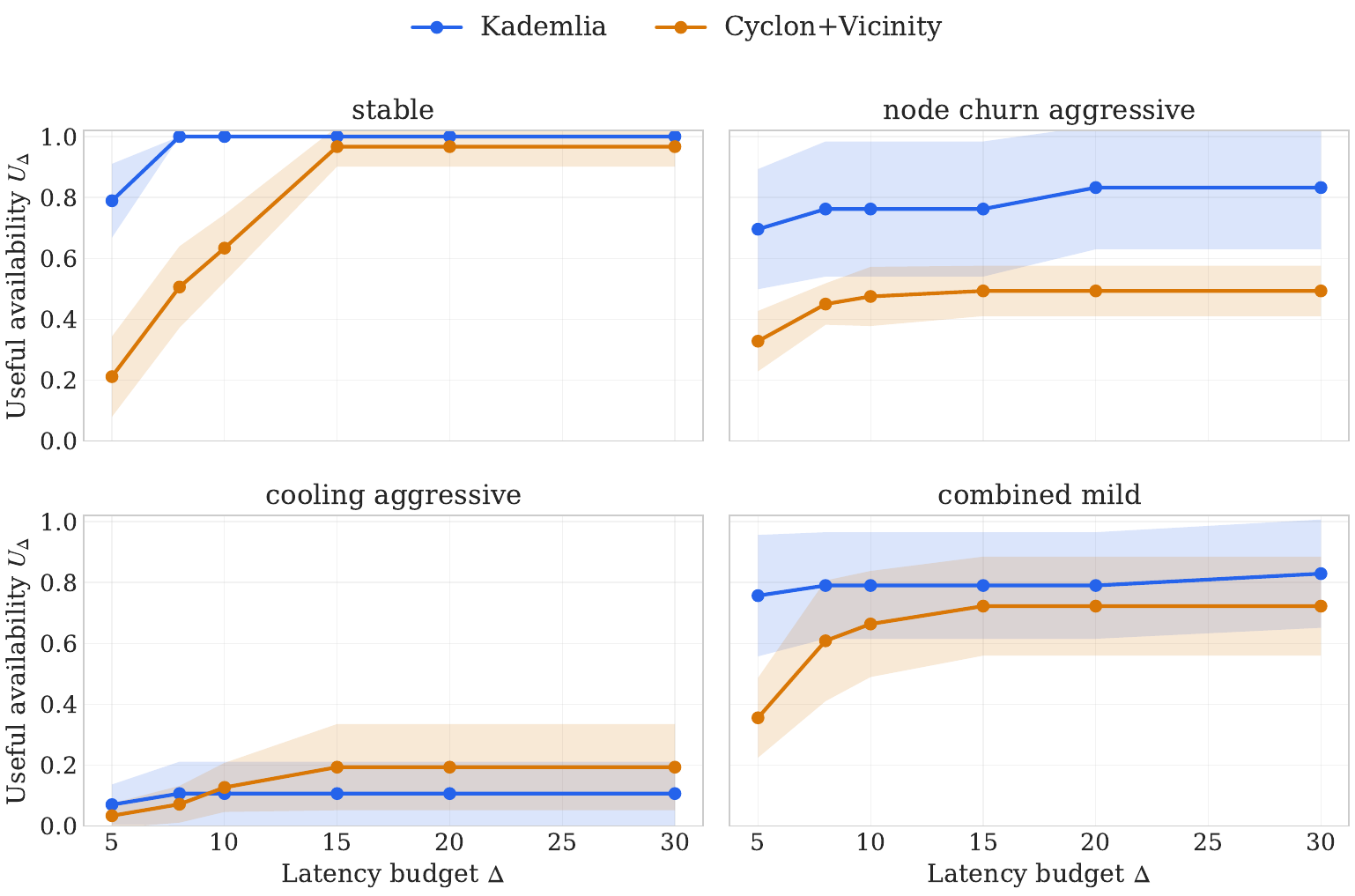}
\caption{Useful availability $\udelta$ vs budget $\Delta$.}
\label{fig:useful-availability}
\end{figure}

\begin{figure}[t]
\centering
\includegraphics[width=\textwidth]{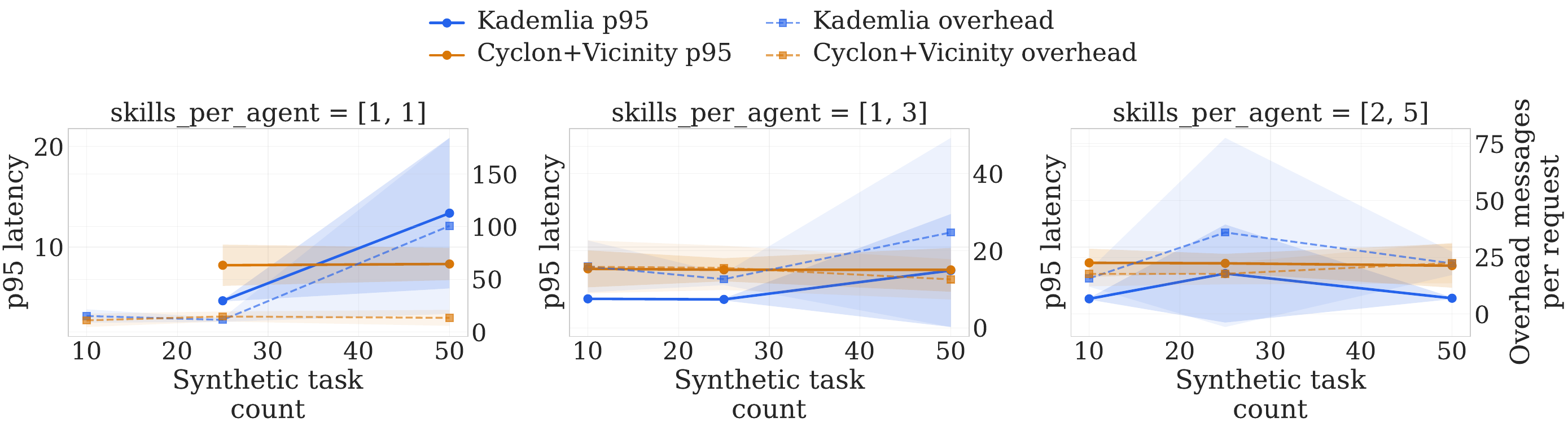}
\caption{p95 latency and overhead from specialists to generalists.}
\label{fig:semantic-density-ablation}
\end{figure}

\subsection{Main findings}

The current implementation supports a coherent reading of the design space on a fully populated core batch. The main multi-host ladders are no longer interpreted only through success and $p95$ latency: Figure~\ref{fig:useful-availability} adds direct useful-availability curves, making it possible to distinguish regimes in which the dominant limitation is routing robustness from regimes in which the dominant limitation is agent readiness. Taken together, Figures~\ref{fig:latency-overhead-tradeoff}--\ref{fig:semantic-density-ablation} support a five-step argument about how the comparison between structured and unstructured discovery should be read.
\noindent\textbf{\textit{Structured efficiency does not imply end-to-end dominance.}}
The stable baseline already shows that a structured overlay does not dominate every visible axis. Figures~\ref{fig:latency-overhead-tradeoff} and~\ref{fig:success-across-regimes} show that \protoK{} holds the stronger success profile in the favorable regime and also achieves lower p95 latency, but the request-plane messaging gap is not dramatic. The main separation lies in maintenance: Figure~\ref{fig:maintenance-cost} shows that \protoK{} pays a much larger background-maintenance bill through index publication and refresh, whereas \protoCV{} pays substantially less through gossip traffic. Structured discovery therefore buys robust lookup, but shifts much more of the cost burden into maintenance.
Low-noise maintenance probes reinforce this interpretation and make the trade-off sharper. More aggressive maintenance can indeed move the operating point, but often at a very steep cost: dense gossip on the unstructured side and aggressive republishing on the structured side both increase background traffic far more than they improve latency or useful discovery quality. The practical lesson is therefore not simply that maintenance matters, but that the two overlay families move along very different cost-quality frontiers, and that ``more maintenance'' is not a universal recovery mechanism.
\noindent\textbf{\textit{Node churn and agent cooling alter the comparison through different mechanisms.}}
Under node churn alone, the structured baseline behaves in a fairly classical way: Figure~\ref{fig:success-across-regimes} shows \protoK{} remaining consistently more robust on success as churn intensifies. At the same time, Figure~\ref{fig:latency-overhead-tradeoff} shows that under aggressive churn the latency advantage can move toward \protoCV{}. Under agent cooling, the interpretation changes again: the success gap compresses and may even flip slightly, while \protoK{} can still retain the lower p95 latency. Node churn therefore stresses routing robustness and maintenance, whereas agent cooling stresses readiness and cold starts.
\noindent\textbf{\textit{Under combined churn the comparison becomes sharply regime dependent.}}
The combined ladder shows why the two-level formulation is necessary. Figures~\ref{fig:success-across-regimes} and~\ref{fig:latency-overhead-tradeoff} together show that the mild combined regime is already a trade-off point rather than an easy structured win: success stays close, while the unstructured overlay is visibly faster. In the moderate rung, \protoK{} regains a modest success edge while the latency difference narrows. In the aggressive rung, both overlays deteriorate sharply, but \protoK{} remains more robust on success. The reduced-scale combined operating point strengthens the same reading: the success gap stays limited, yet \protoCV{} remains distinctly faster. The combined ladder therefore suggests not a clean inversion of winners, but a persistent trade-off in which structured discovery buys more robustness while unstructured discovery can still buy better latency in nearby operating conditions.
\noindent\textbf{\textit{Useful availability reveals when routing quality still converts into usable service.}}
Figure~\ref{fig:useful-availability} shows $\udelta$ for four regimes. \protoK{} dominates at tight budgets in stable and node-churn regimes. Under aggressive cooling, \protoK{} flattens while \protoCV{} improves with looser budgets, confirming readiness dominates routing.
\noindent\textbf{\textit{Semantic sparsity matters mainly when agents remain highly specialized.}}
Figure~\ref{fig:semantic-density-ablation} shows that the clearest semantic effect appears in the specialist setting. With one skill per agent, the two overlays remain visibly separated as task count grows, whereas with more generalist agents [1,3] or [2,5] the p95-latency and overhead curves move much closer together. The main lesson is therefore not that semantic density reshapes every operating point, but that it matters most when capability placement is highly concentrated.
Targeted diagnostic probes also sharpen the causal reading of the instability results. In those probes, when pressure is focused on host-belief freshness, the structured overlay recovers much of its expected strength; when pressure is focused on routing freshness, it remains brittle despite substantially higher traffic. This does not replace the broader regime comparison, but it strongly suggests that the most severe failures under combined instability are more consistent with stale routing than with a generic lack of host-belief freshness.
Taken together, these findings support the central claim of the paper: structured and unstructured discovery cannot be ranked by a single scalar notion of ``better.'' In our evaluation, \protoK{} is systematically stronger on success and often on useful availability, but the size of that advantage varies substantially by regime, while \protoCV{} remains competitive or superior on latency in several operating points. The newer maintenance and staleness probes do not overturn that conclusion; instead, they make it more precise by showing where the traffic budget is spent and which failure mechanism appears to dominate once the system becomes unstable.

\subsection{Targeted Maintenance and Staleness Probes}

\begin{figure}[t]
\centering
\includegraphics[width=\textwidth]{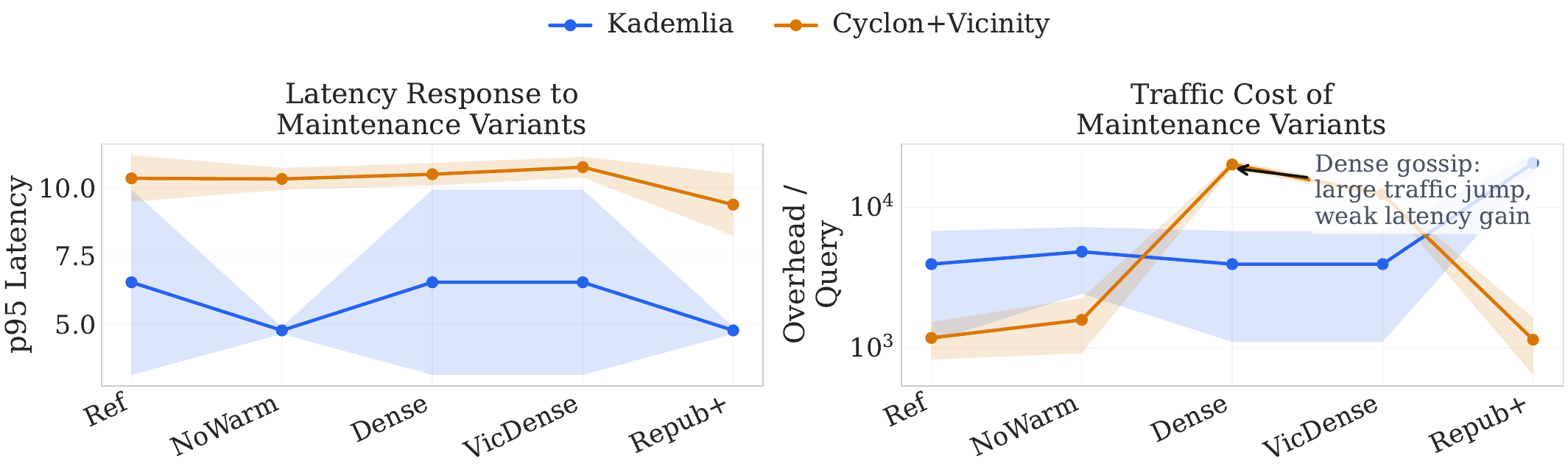}
\caption{Targeted maintenance probe with 95\% confidence intervals.}
\label{fig:maintenance-probe-targeted}
\end{figure}

\begin{figure}[t]
\centering
\includegraphics[width=\textwidth]{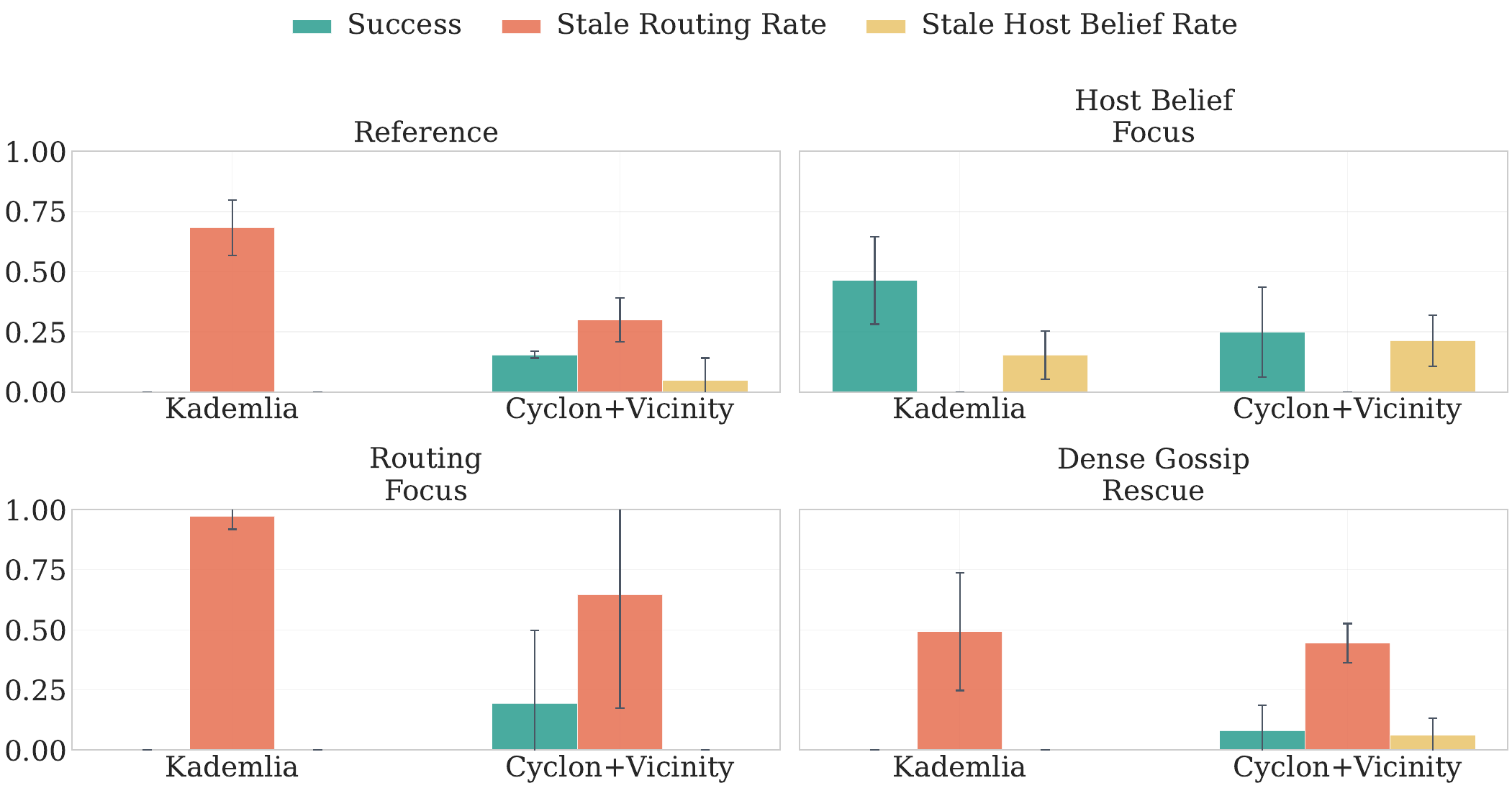}
\caption{Targeted staleness diagnostics with 95\% confidence intervals.}
\label{fig:stale-diagnostics-targeted}
\end{figure}

The targeted rerun campaign complements the main regime ladders with two
probes. These probes are narrower than the full study, but they make the
current interpretation more explicit by showing how maintenance intensity and
staleness mechanisms reshape the comparison.
The last two figures are therefore not additional operating points from the
main ladder. Figure~\ref{fig:maintenance-probe-targeted} isolates warmup and
maintenance intensity under low-noise conditions, whereas
Figure~\ref{fig:stale-diagnostics-targeted} separates stale-routing effects
from stale host-belief effects. Table~\ref{tab:targeted-probes} summarizes the
controlled variants used in these probes.

\begin{table}[t]
\centering
\small
\setlength{\tabcolsep}{3pt}
\caption{Targeted probe configurations used in Figures~\ref{fig:maintenance-probe-targeted} and~\ref{fig:stale-diagnostics-targeted}.}
\label{tab:targeted-probes}
\begin{tabular}{llp{0.58\textwidth}}
\toprule
Probe & Variant & Distinguishing parameters \\
\midrule
Maintenance & Reference & 1024 agents, 512 hosts, no churn, warmup 9 \\
Maintenance & No warmup & bootstrap only, warmup 0 \\
Maintenance & Dense gossip & Cyclon shuffle 1, Vicinity shuffle 1, forward $k=2$ \\
Maintenance & Vicinity dense & Cyclon shuffle 20, Vicinity shuffle 1, forward $k=2$ \\
Maintenance & Republish+ & \protoK{} republish period 1 \\
Staleness & Reference & 512 ag., 256 hosts, downtime 20, lifecycle 18/6/2 \\
Staleness & Host-belief focus & no node churn, lifecycle 10/10/4 \\
Staleness & Routing focus & session 80, downtime 25, lifecycle disabled \\
Staleness & Dense-gossip rescue & reference staleness with Cyclon/Vicinity shuffle 1 \\
\bottomrule
\end{tabular}
\end{table}

\noindent\textbf{\textit{Maintenance changes cost much faster than it changes
quality.}}
Figure~\ref{fig:maintenance-probe-targeted} makes the maintenance story more
concrete than the aggregate regime ladder alone. In the targeted probe, dense
gossip on the unstructured side increases overhead by roughly an order of
magnitude, yet yields little or no convincing improvement in latency. On the
structured side, aggressive republishing likewise expands the traffic bill far
more than it improves an already strong operating point. This is useful because
it rules out a simplistic reading of the comparison: the central issue is not
that one family merely needs ``more maintenance,'' but that the two families
move along sharply different cost-quality frontiers.

\noindent\textbf{\textit{The sharpest unstable failures are more consistent
with stale routing than with generic host-belief loss.}}
Figure~\ref{fig:stale-diagnostics-targeted} strengthens the causal interpretation
of the instability results. In the reference diagnostic regime, the structured
overlay can collapse almost entirely while \protoCV{} still achieves partial
discovery. When the probe is shifted toward host-belief pressure, however,
\protoK{} recovers much of its expected strength;
when the probe is shifted toward routing pressure, it remains brittle despite
clear mitigation effort. The dense-gossip rescue case reinforces the same
point: even a more aggressive unstructured rescue does not restore a convincing
quality advantage. These probes therefore suggest that the most severe
breakdowns under combined instability are better explained by stale routing than
by a generic lack of host-belief freshness.

\section{Conclusions and Next Steps}

We treat decentralized discovery as a problem under two-level churn. By separating node instability from agent readiness, we show structured and unstructured overlays occupy different regimes rather than yielding a single winner.
Discovery must be evaluated on lookup efficiency, robustness, and readiness together. Node churn and agent cooling interact nonlinearly: structured overlays usually succeed more often, while gossip overlays are cheaper to maintain and often lower latency. Thus $\udelta$ is the key metric for usable service~\cite{dean2013tail}, not a secondary concern.
Three results follow. First, useful availability $\udelta$ — not success rate alone — reveals when routing robustness yields timely service. Under stable and node-churn regimes, \protoK{} turns its success advantage into near-complete $\udelta$ even with tight latency; under aggressive cooling this fails because cold starts dominate precise routing. Second, maintenance and failure modes diverge: structured overlays pay steady maintenance for fresh indices, while gossip overlays pay on demand. Targeted probes show most combined-instability failures stem from stale routing, not stale host-belief: \protoK{} recovers when host-belief is stressed but stays fragile when routing is stressed. Third, semantic sparsity matters only under extreme specialization: once agents have moderately broad skills, both families achieve near-full success and the trade-off reduces to latency versus overhead.
Overlay choice should follow churn, SLOs, and not just asymptotic lookup cost. Stringent latency and robustness needs favor structured designs; cost-sensitive settings that tolerate graceful degradation favor gossip. 
Two directions emerge. First, hybrids that use structured lookup in stable periods but fall back to unstructured repair when routing tables decay. Second, request-aware lifecycle management that aligns agent warmness with overlay health and predicted demand, turning $\udelta$ from a diagnostic into a control objective.
\bibliographystyle{splncs04}
\bibliography{biblio}

\end{document}